\documentclass[conference]{IEEEtran}
\IEEEoverridecommandlockouts

\usepackage[utf8]{inputenc}
\usepackage[T1]{fontenc}
\usepackage{microtype,newtxtext,newtxmath} 
\usepackage{mathtools,gensymb,url} 
\usepackage{subcaption}%
\usepackage[font=footnotesize,textfont=footnotesize]{subcaption}
\usepackage{longtable,booktabs,array}
\usepackage[backend=biber, sorting=none, style=ieee, maxnames=5]{biblatex}
\usepackage{acronym}
\usepackage{graphicx}
\usepackage{hyperref}
\usepackage{multirow}
\hypersetup{colorlinks=true, breaklinks=true, 
  urlcolor=blue, linkcolor=blue,anchorcolor=blue,citecolor=blue,
  hypertexnames=true, final=true, 
  pdfpagemode = UseNone, 
  pdfauthor = {},
  pdftitle = {},   
  pdfsubject = {},
  pdfkeywords = {}
}

\graphicspath{{img/} {images/} {./}}
\DeclareGraphicsExtensions{.pdf,.png,.jpg,.mps}

\IEEEdisplaynontitleabstractindextext

\DeclareSourcemap{
  \maps[datatype=bibtex]{
    \map[overwrite=true]{
      \step[fieldset=url, null]
      \step[fieldset=doi, null]
      \step[fieldset=eprint, null]
       \step[fieldset=month, null]  
    }
  }
}

\acrodef{EEG}[EEG]{Electroencephalography}
\acrodef{EMG}[EMG]{Electromyography}
\acrodef{HFO}[HFO]{High Frequency Oscillation}
\acrodef{ECG}[ECG]{Electrocardiography}
\acrodef{SNN}[SNN]{spiking neural network}
\acrodef{DPI}[DPI]{Differential Pair Integrator}
\acrodef{ADM}[ADM]{Asynchronous Delta Modulation}
\acrodef{RMSE}[RMSE]{Root Mean Square Error}
\acrodef{E-I}[E-I]{excitatory-inhibitory} 
\acrodef{SVM}[SVM]{Support Vector Machine}
\acrodef{HR}[HR]{Heart Rate}

\acrodef{AdEx InF}[AdEx I\&F]{adaptive exponential integrate-and-fire model}
\acrodef{CAM}[CAM]{Content Addressable Memory}
\acrodef{AER}[AER]{Address-Event Representation}
\acrodef{bpm}[bpm]{beats per minute}
\acrodef{LIF}[LIF]{Leaky Integrate-and-Fire}
\acrodef{PPG}[PPG]{photoplethysmograms}
\acrodef{ECG}[ECG]{electrocardiogram} 
\acrodef{NSM}[NSM]{Neural State Machine}
\acrodef{EMG}[EMG]{Electromyography}
\acrodef{HFO}[HFO]{High Frequency Oscillation}
\acrodef{HR}[HR]{Heart Rate}
\acrodef{WTA}[WTA]{winner-take-all}
\acrodef{EI}[EI]{Excitatory-Inhibitory}

\begin{document}
\title{Neuromorphic Heart Rate Monitors: Neural State Machines for Monotonic Change Detection}

\author{\IEEEauthorblockN{
Alessio Carpegna\IEEEauthorrefmark{1}, 
Chiara De Luca\IEEEauthorrefmark{2}\IEEEauthorrefmark{3}, 
Federico Emanuele Pozzi\IEEEauthorrefmark{4}, 
Alessandro Savino\IEEEauthorrefmark{1},
Stefano Di Carlo\IEEEauthorrefmark{1},\\
Giacomo Indiveri\IEEEauthorrefmark{2}, 
Elisa Donati\IEEEauthorrefmark{2}%
}
\vspace{0.2cm}

\IEEEauthorblockA{\IEEEauthorrefmark{1}Department of Control and Computer Engineering, Politecnico di Torino, Torino, Italy}
\IEEEauthorblockA{\IEEEauthorrefmark{2}Institute of Neuroinformatics, University of Zurich and ETH Zurich, Zurich, Switzerland\\ }
\IEEEauthorblockA{\IEEEauthorrefmark{3}Digital Society Initiative, University of Zurich, Zurich, Switzerland\\}
\IEEEauthorblockA{\IEEEauthorrefmark{4}Neurology Department, Fondazione IRCCS San Gerardo Dei Tintori, Monza, Italy\\
\vspace{-1.3cm}}

\thanks{This paper has received funding from: The NEUROPULS project in the European Union’s Horizon Europe research and innovation programme under grant agreement No. 101070238; Bridge Fellowship founded by the Digital Society Initiative at University of Zurich (grant no.G-95017-01-12).  We also thank the University of Zurich for supporting this project.}
}
\maketitle

\begin{abstract}
Detecting monotonic changes in heart rate (HR) is crucial for early identification of cardiac conditions and health management. This is particularly important for dementia patients, where HR trends can signal stress or agitation. Developing wearable technologies that can perform always-on monitoring of HRs is essential to effectively detect slow changes over extended periods of time. 
However, designing compact electronic circuits that can monitor and process bio-signals continuously, and that can operate in a low-power regime to ensure long-lasting performance, is still an open challenge.
Neuromorphic technology offers an energy-efficient solution for real-time health monitoring. We propose a neuromorphic implementation of a Neural State Machine (NSM) network to encode different health states and switch between them based on the input stimuli. Our focus is on detecting monotonic state switches in electrocardiogram data to identify progressive HR increases. This innovative approach promises significant advancements in continuous health monitoring and management.
\end{abstract}


\section{Introduction}
\label{sec:intro}
Detecting and quantifying \ac{HR} has emerged as a critical tool for identifying potential pathologies and providing valuable insights into cardiovascular health~\cite{Rajendra_etal2006}. Among variable behaviors, monotonic HR changes indicate unidirectional trends, either increasing or decreasing, in average HR over time.
In non-clinical settings, such as general well-being and athletic training, tracking monotonic changes in \ac{HR} is essential for evaluating physical fitness and recovery rates~\cite{Zhu_etal22}. 
In clinical settings, monitoring monotonic changes in \ac{HR} is crucial for medical diagnosis and patient monitoring. A consistent monotonic increase or decrease in heart rate can be an early indicator of cardiac conditions such as arrhythmia, bradycardia, or tachycardia. 
Early detection of these trends enables timely medical intervention, preventing more severe complications and improving patient outcomes~\cite{Heidenreich_etal11}.

Understanding monotonic increases in \ac{HR} is also particularly important in patients with dementia, as it can help detect physiological stress or discomfort that could precede or accompany agitation states~\cite{Liu_etal23, Davidoff_etal23}. By monitoring trends in dementia care and detecting early signs of agitation, healthcare providers and caregivers can intervene more promptly and effectively. This may potentially reduce the severity and frequency of agitation episodes, thereby improving the quality of care and the quality of life for dementia patients and reducing the burden on caregivers~\cite{Caspar_etal17}.

Due to the importance and, at the same time, the difficulty of detecting these changes for human interpreters, an always-on system capable of continuously monitoring and detecting changes in average \acp{HR} can be extremely impactful. Always-on devices for health monitoring must meet stringent requirements, including low power consumption and real-time processing. For patients with dementia, these devices must operate for extended periods without frequent recharging, as subjects might not remember to charge the device regularly. Therefore, a low-power device that can be used as a ``wear and forget'' system is essential to ensure effective health management. 

To this end, neuromorphic technology offers a compelling solution, providing energy-efficient and reliable processing capabilities~\cite{Chicca_etal14}. Sensory-processing systems built using mixed-signal neuromorphic circuits are well-suited to the demands of continuous health monitoring~\cite{Donati_Indiveri23}. Example solutions have already been successfully applied in a wide range of wearable applications, such as \ac{ECG} anomaly detection~\cite{Bauer_etal19, Das_etal18a}, \ac{HFO} detection~\cite{Sharifshazileh_etal21}, and \ac{EMG} decoding~\cite{Ma_etal20, Vitale_etal22}.

\begin{figure*}
    \centering
    \includegraphics[width=0.8\textwidth]{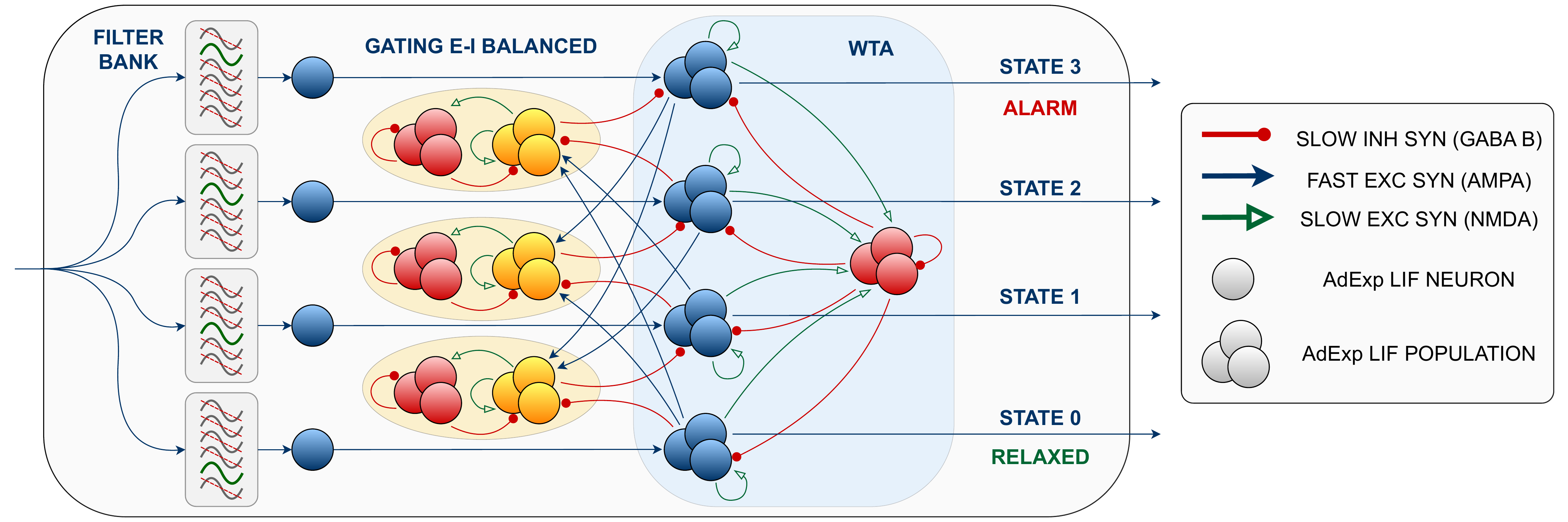}
    \caption{Monotonic \ac{NSM}. The input signal is filtered through four $4^{th}$ order Butterworth bandpass filters. Each filtered component is converted into spikes through a \ac{AdEx InF} neuron (blue). The input spikes are fed into populations of \ac{AdEx InF} neurons (blue), encoding different states of the network, interconnected in a \ac{WTA} configuration via a common inhibitory population (red). States are connected to gating populations  (yellow) to implement the monotonic computation. These are organized in a \ac{EI}-balanced configuration with an inhibitory population (red) limiting the overall activity. }
    \label{fig:monotonic_nsm}
\end{figure*}

In this paper, we present for the first time a neuromorphic implementation of an on-line signal processing system to specifically detect monotonic changes in average \acp{HR} over a long period of time. We deploy computational primitives of analog neuron circuits, such as recurrent neural network models of finite state machines (i.e., \acp{NSM})~\cite{Neftci_etal13,Liang_Indiveri19} that can switch between states, each encoding different average \ac{HR} conditions, independent of the time elapsed between state changes. Here, we focus on a monotonic state switch, tested on \ac{ECG}, to detect a progressive \ac{HR} increase. We validate our model's ability to track \ac{HR} changes during activities (i.e., walking, cycling) by testing it on a dataset with varying patterns. We demonstrate how the model accurately follows monotonic changes (steady increase or decrease) while remaining inactive for non-monotonic signals. We further show how the robust computational properties of \acp{NSM} allow the mixed-signal neuromorphic processor to produce accurate and reliable results. The low power consumption, $90 \mu W$, and real-time processing features of these neuromorphic circuits make them ideal candidates for building continuous, long-term health monitoring devices in both clinical and non-clinical settings.

\section{Materials and Methods}
\label{sec:matmet}

\subsection{Neuromorphic Hardware}
\label{ssec:chip}
The neuromorphic processor used in this study is the DYNAP-SE chip~\cite{Moradi_etal18}. It is a custom-designed asynchronous mixed-signal processor that features analog spiking neurons and synapses that mimic the biophysical properties of their biological counterparts in real-time. The chip comprises four cores, each containing 256 \ac{AdEx InF} neurons. Each synapse can be configured as one of four types: slow/fast and inhibitory/excitatory. Each neuron includes a \ac{CAM} block with 64 addresses, representing the pre-synaptic neurons to which it is connected. Digital peripheral asynchronous input/output logic circuits receive and transmit spikes via the \ac{AER} communication protocol~\cite{Deiss_etal98}. In this system, each neuron is assigned a unique address encoded as a digital word, which is transmitted using asynchronous digital circuits as soon as an event is generated. The chip features a fully asynchronous inter-core and inter-chip routing architecture, allowing flexible connectivity with microsecond precision even under heavy system loads.


\subsection{Dataset and signal processing}
\label{ssec:signal}
To test our model, we selected an available dataset that includes left wrist \acp{PPG} and chest \ac{ECG} recordings taken while participants used an indoor treadmill and exercise bike, accompanied by simultaneous motion data from accelerometers and gyroscopes~\cite{Jarchi_etal16}. Participants performed various exercises, such as walking, light jogging/running on a treadmill, and pedaling at low and high resistance, each for up to 10 minutes. The dataset includes records from 8 participants, with most participants spending 4 to 6 minutes per activity. The signals were sampled at 256 Hz, and the \ac{ECG} records were processed with a 50 Hz notch filter to remove mains interference.

An energy-based approach was devised for signal-to-spike conversion~\cite{Narayanan_etal23}. The proposed method comprises two stages: bandpass filtering and \ac{LIF} neurons~\cite{Zanghieri_etal23, Sava_etal23, Zanghieri_etal24}.
Each input channel was processed through a series of bandpass filters. We used four bands: 60-82 (\#0), 82-105 (\#1), 105-128 (\#2), and 128-150 (\#3) \ac{bpm}, employing fourth-order Butterworth filters. This covers a reasonable range of \ac{HR} variation, spanning from a relaxed state to a tachycardiac one. Afterward, the signal was full-wave rectified and injected as a time-varying current into a simple \ac{LIF} neuron. 

\subsection{Network on chip}
\label{ssec:network}

    \begin{figure*}[t]
        \centering
        \includegraphics[width=0.8\textwidth]{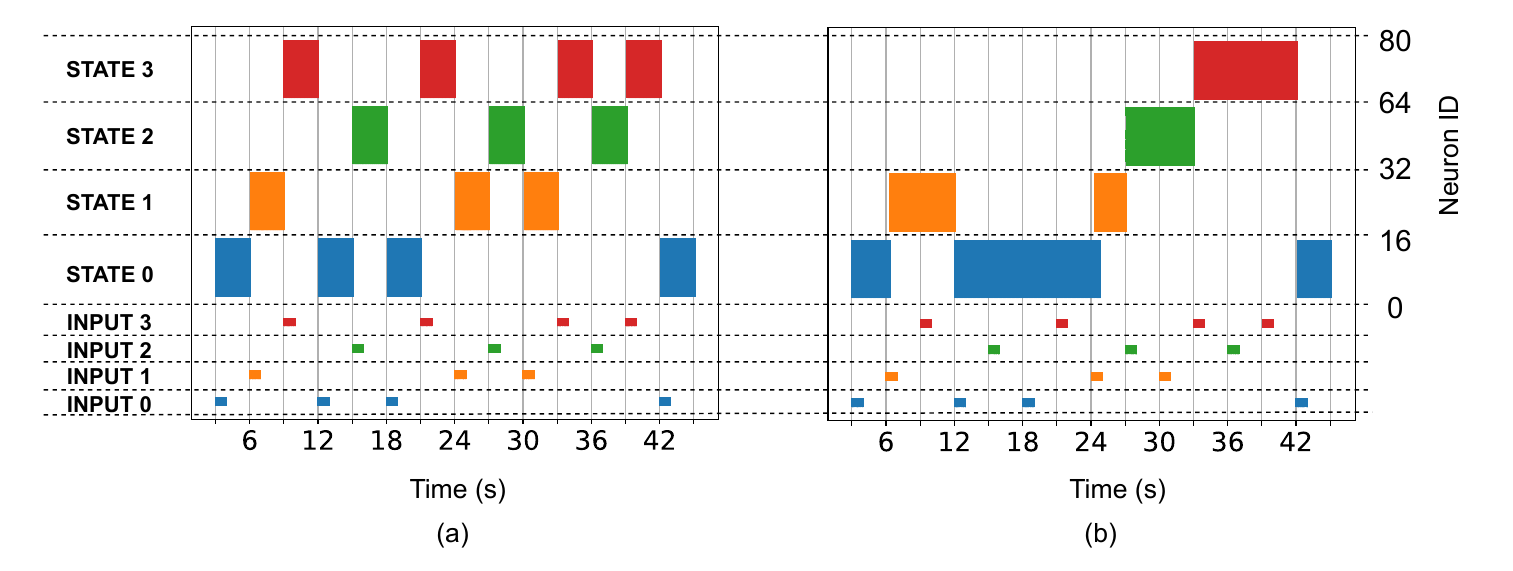}
        \caption{Network behavior when stimulated with 50Hz Poissonian sequences testing all the possible transitions: (a) non-monotonic \ac{WTA} dynamics; (b) monotonic \ac{NSM} response}
        \label{fig:poisson}
    \end{figure*}

Figure~\ref{fig:monotonic_nsm} shows the designed \ac{NSM}. The architecture is based on \ac{EI}-balanced populations of \ac{AdEx InF} neurons. These populations can maintain sustained activity for extended periods, much larger than the time constants of the neurons themselves. This allows the network to implement a working memory capable of processing signals that change slowly compared to the time scales used within the chip.

The core of the network is a set of four \ac{EI}-balanced populations, organized in a \acf{WTA} architecture. The shared inhibition population (red) encodes the state of the network by sustaining the activity of a specific population while silencing all the others. This guarantees that when receiving the input signal, such as an \ac{ECG} recording, only the population corresponding to the most active frequency range is activated. 

A second set of \ac{EI}-balanced populations, named Gating \ac{EI}-balanced populations, is used to control the transition between different states. The goal is to target a progressive increase in \ac{HR}, making the system robust to short fluctuations or temporary \acp{bpm} decreases, following only the average trend of the input. 
The mechanism of dis-inhibition~\cite{Liang_Indiveri19} is exploited to ensure the network switches only to increasing states. The gating populations continuously inhibit the inactive elements of the \ac{WTA} network, making them insensitive to any input stimulus. When a state is activated, it turns off (inhibits) the gating population of the subsequent state. This leads to the dis-inhibition of the next state, which makes it sensitive to the input. At the same time, it activates all the gating populations of previous states and the states following the next. This guarantees that the network follows only monotonic transitions between states and does not require precise tuning of the populations and connection synapses to make the system work. What matters is (i) that populations are sufficiently stable to maintain a sustained activity and (ii) that the inhibition between the gating populations and the \ac{WTA} states is strong enough to silence any spiking activity completely. Finally, state 0 has no gating populations connected to it. This allows it to be used as a reset state: if the monotonic increase is only partial, and the \ac{HR} returns to the relaxation range before reaching the alarming threshold, the network restarts, waiting for a new ramp-up in the input.

\section{Experimental Results}
\label{sec:results}

To obtain reliable computation in the \ac{NSM} while minimizing the overall network size, we used 16 neurons per population, with a total requirement of 176 neurons. Table~\ref{tab:connections} shows the average connection probabilities for different populations. The resulting network is compact, fitting well within a single core of  the target chip~\cite{Moradi_etal18}. Each population exhibits an average firing rate of approximately 50Hz. The total power consumption can be estimated by integrating the various contributions required for spike generation and communication~\cite{Risi_etal20, Zhao_etal23}. The obtained average power consumption is around $90 \mu W$, indicating that our network successfully balances stable, sustained activity with low power consumption.

\begin{table}[ht]
        \caption{Connection types and average probabilities}
        \label{tab:connections}
        \centering
        \begin{tabular}{|c|c|c|c|}

            \cline{2-4}
            \multicolumn{1}{c|}{}                              &
            \textbf{Connection}                               &
            \textbf{Synapse type}                                 &
            \textbf{Probability}                              \\

            \hline

            \multirow{4}{*}{\rotatebox[origin=c]{0}{\textbf{WTA}}}      &
            state -> inh                                         &
            NMDA                                         &
            60\%                                                \\

            \cline{2-4}

            {}                                                  &
            inh -> state                                         &
            GABA B                                         &
            60\%                                                \\
            
            \cline{2-4}

            {}                                                  &
            state -> state                                         &
            NMDA                                         &
            83\%                                                \\

            \cline{2-4}

            {}                                                  &
            inh -> inh                                         &
            GABA B                                         &
            20\%                                                \\
            
            \hline

            \multirow{4}{*}{\rotatebox[origin=c]{0}{\textbf{GATING}}}      &
            gate -> inh                                         &
            NMDA                                         &
            30\%                                                \\

            \cline{2-4}

            {}                                                  &
            inh -> gate                                         &
            GABA B                                         &
            30\%                                                \\
            
            \cline{2-4}

            {}                                                  &
            gate -> gate                                         &
            NMDA                                         &
            50\%                                                \\

            \cline{2-4}

            {}                                                  &
            inh -> inh                                         &
            GABA B                                         &
            50\%                                                \\

            \hline

            \multirow{3}{*}{\rotatebox[origin=c]{0}{\textbf{MONOTONIC}}}      &
            gate -> state                                         &
            GABA B                                         &
            100\%                                                \\

            \cline{2-4}

            {}                                                  &
            state -> gate                                         &
            GABA B                                         &
            100\%                                                \\
            
            \cline{2-4}

            {}                                                  &
            state -> gate                                         &
            AMPA                                         &
            100\%                                                \\
            
            \hline

            \textbf{INPUT}                                                  &
            lif -> state                                         &
            AMPA                                         &
            100\%                                                \\
            
            \hline
            
        \end{tabular}
    \end{table}

        \begin{figure}[ht]
        \centering
        \includegraphics[width=0.9\columnwidth]{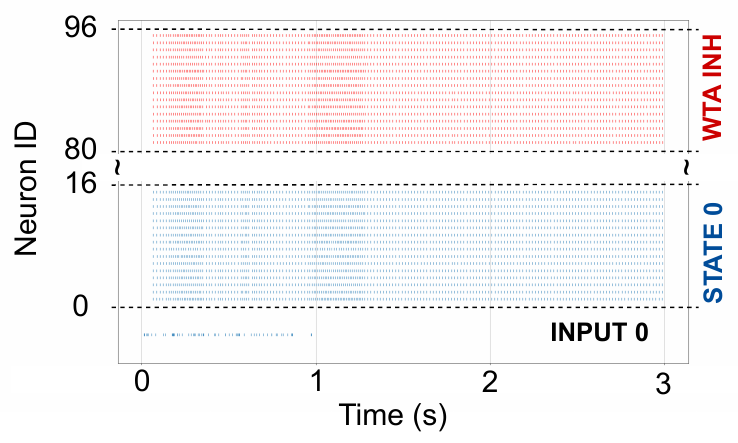}
        \caption{Stable sustained activity of one of the \ac{EI}-balanced primitives included in the networks}
        \label{fig:stable}
    \end{figure}

    \begin{figure*}[ht]
        \centering
        \includegraphics[width=0.9\textwidth]{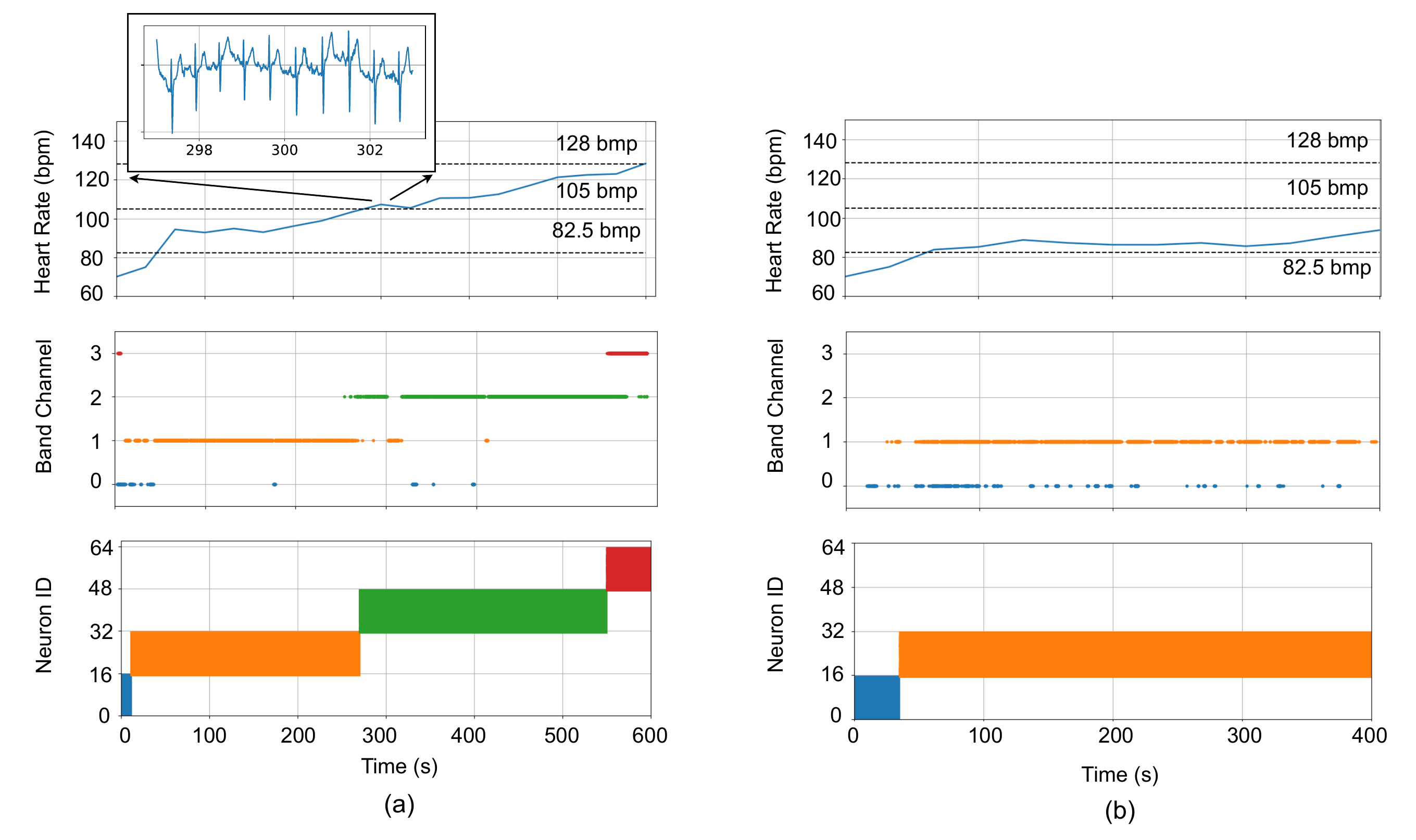}
        \caption{Response of the network when stimulated with a real \ac{ECG} signal: (a) intense 10 minutes bike session; (b) 6 minutes and 40 seconds of walk}
        \label{fig:ecg_complete}
    \end{figure*}

The behavior of the hardware system was first tested using control stimuli produced as 50Hz Poisson input spike trains. Figure~\ref{fig:stable} shows the sustained activity of one \ac{EI}-balanced population stimulated for one second. As shown, the network can keep a stable activity even when the input is removed. 

\subsection{WTA network: non-monotonic state transitions}
\label{ssec:non_monotonic_res}
As a second test, we evaluated the dynamics of the \ac{WTA} network. Figure~\ref{fig:poisson}(a) shows the response of the network when stimulated with Poisson spike trains through a protocol covering all the possible transitions. In this case, each state is activated when receiving the corresponding input, regardless of the order of arrival. Once stimulated, the population becomes active, it suppresses the activity of all other populations, and stays active until a new input is provided to a different population.

\subsection{NSM network: monotonic state transitions}
\label{ssec:monotonic_res}
When adding the gating connections, the response of the network becomes the one shown in Fig.~\ref{fig:poisson}(b). The stimulation protocol is the same as in Section~\ref{ssec:non_monotonic_res}. In this case, however, the network switches state only (i) towards  a higher state, following the expected monotonic behavior, or (ii) towards state 0, which behaves as a reset state for the network, as mentioned in Section~\ref{ssec:network} 

\subsection{Network dynamics on real ECG signal}
\label{ssec:ecg_res}
Finally, we evaluated the network performance using real \ac{ECG} signals. Here, we demonstrate its robust and coherent ability to monitor and detect monotonic increases in \acp{HR}. To achieve this, we fed the network pre-processed ECG recordings from the dataset described in Section~\ref{ssec:signal}. Specifically, we tested the network under two different conditions: during an intense physical activity, namely cycling, lasting approximately 10 minutes (Fig.~\ref{fig:ecg_complete}(a)), and during a walking session of around 6 minutes (Fig.~\ref{fig:ecg_complete}(b)). The first scenario correctly detects a complete ramp-up of the \ac{HR}, leading to the activation of the alarming state. These results show how the network is robust to spurious transitions: at time zero, the stimulation on input~3, caused by noise in the recording upper band, is ignored since state~3 is completely inhibited by its gating population. The same effect can be observed between 300 and 400 seconds with the network ignoring noisy stimuli from input~1. Note that also the weak stimuli coming from input 0 are ignored both in state~1 and~2, despite the absence of a gating population in this case. This shows that the \ac{WTA} dynamics by themselves are robust to spurious transitions and a complete relaxation is required to reset the network, thus restarting the monotonic increase. This is even more evident in the walking session: in this case, the ramp-up is only partial, given the lower effort required. The network remains stable in state~1, waiting for a further increase in the \ac{HR} and ignoring the noise on input~0. 





\section{Conclusion}
\label{sec:conclusion}
Our results show that a small and simple network, implemented with mixed-signal analog/digital neuromorphic circuits, can reliably monitor a monotonic \ac{HR} trends over extended periods, paving the way for always-on health monitoring systems that are both efficient and long-lasting. The low power consumption of the neuromorphic circuits enables continuous operation for extended amounts of time, making it ideal for wearable devices for health monitoring. Future research will focus on replacing \ac{ECG} with \ac{PPG} signals measured at the wrist and addressing the challenges of noise and movement artifacts. Additionally, we will transition from healthy subjects to patients affected by neuropathologies, such as dementia, which impact \ac{HR} over long periods, despite the absence of cardiac pathology.

This technology promises significant improvements in patient care and health management, especially in scenarios requiring constant monitoring and rapid response, thus enhancing overall quality of life.

\printbibliography

@Article{Bauer_etal19,
author		= {Bauer, Felix and Muir, Dylan and Indiveri, Giacomo},
title		= {Real-time ultra-low power {ECG} anomaly detection using an
		  event-driven neuromorphic processor},
journal		= {Biomedical Circuits and Systems, {IEEE} Transactions on},
year		= {2019},
month		= dec,
volume		= {13},
number		= {6},
pages		= {1575--1582},
doi		= {10.1109/TBCAS.2019.2953001}
}

@Article{Chicca_etal14,
author		= {E. Chicca and F. Stefanini and C. Bartolozzi and G.
		  Indiveri},
title		= {Neuromorphic electronic circuits for building autonomous
		  cognitive systems},
journal		= {Proceedings of the {IEEE}},
year		= {2014},
month		= sep,
volume		= {102},
number		= {9},
pages		= {1367--1388},
issn		= {0018-9219},
doi		= {10.1109/JPROC.2014.2313954}
}

@Article{Das_etal18a,
author		= {Anup Das and Paruthi Pradhapan and Willemijn Groenendaal
		  and Prathyusha Adiraju and Raj Thilak Rajan and Francky
		  Catthoor and Siebren Schaafsma and Jeffrey L. Krichmar and
		  Nikil D. Dutt and Chris Van Hoof},
title		= {Unsupervised heart-rate estimation in wearables with
		  Liquid states and a probabilistic readout},
journal		= {Neural networks},
year		= {2018},
volume		= {99},
pages		= { 134--147 },
doi		= {10.1016/j.neunet.2017.12.015}
}

@InCollection{Deiss_etal98,
author		= {S.R. Deiss and R.J. Douglas and A.M. Whatley},
title		= {A Pulse-Coded {C}ommunications Infrastructure for
		  Neuromorphic Systems},
booktitle	= {Pulsed Neural Networks},
chapter		= {6},
year		= {1998},
pages		= {157--78},
publisher	= {MIT Press},
editor		= {W. Maass and C.M. Bishop},
doi		= {10.7551/mitpress/5704.003.0011}
}

@Article{Liang_Indiveri19,
author		= {D. Liang and G. Indiveri},
title		= {A neuromorphic computational primitive for robust
		  context-dependent decision making and context-dependent
		  stochastic computation},
journal		= {{IEEE} Transactions on Circuits and Systems II: Express
		  Briefs},
year		= {2019},
volume		= {66},
number		= {5},
pages		= {843--847},
doi		= {10.1109/TCSII.2019.2907848}
}

@InProceedings{Ma_etal20,
author		= {Ma, Yongqiang and Donati, Elisa and Chen, Badong and Ren,
		  Pengju and Zheng, Nanning and Indiveri, Giacomo},
title		= {Neuromorphic Implementation of a Recurrent Neural Network
		  for {EMG} Classification},
booktitle	= {International Conference on Artificial Intelligence
		  Circuits and Systems ({AICAS}),2020},
year		= {2020},
pages		= {69--73},
organization	= {IEEE},
doi		= {10.1109/AICAS48895.2020.9073810}
}

@Article{Moradi_etal18,
author		= {Moradi, S. and Qiao, N. and Stefanini, F. and Indiveri,
		  G.},
title		= {A Scalable Multicore Architecture With Heterogeneous
		  Memory Structures for Dynamic Neuromorphic Asynchronous
		  Processors ({DYNAPs})},
journal		= {Biomedical Circuits and Systems, {IEEE} Transactions on},
year		= {2018},
month		= feb,
volume		= {12},
number		= {1},
pages		= {106--122},
doi		= {10.1109/TBCAS.2017.2759700}
}

@Article{Neftci_etal13,
author		= {E. Neftci and J. Binas and U. Rutishauser and E. Chicca
		  and G. Indiveri and R. Douglas},
title		= {Synthesizing cognition in neuromorphic electronic
		  systems},
journal		= {Proceedings of the National Academy of Sciences},
year		= {2013},
volume		= {110},
number		= {37},
pages		= {E3468--E3476}
}

@Article{Risi_etal20,
author		= {Risi, Nicoletta and Aimar, Alessandro and Donati, Elisa
		  and Solinas, Sergio and Indiveri, Giacomo},
title		= {A Spike-Based Neuromorphic Architecture of Stereo Vision},
journal		= {Frontiers in Neurorobotics},
year		= {2020},
volume		= {14},
pages		= {93},
issn		= {1662-5218},
doi		= {10.3389/fnbot.2020.568283},
url		= {https://www.frontiersin.org/article/10.3389/fnbot.2020.568283}
}

@Article{Sharifshazileh_etal21,
author		= {Sharifshazileh, Mohammadali and Burelo, Karla and
		  Sarnthein, Johannes and Indiveri, Giacomo},
title		= {An electronic neuromorphic system for real-time detection
		  of High Frequency Oscillations ({HFOs}) in intracranial
		  {EEG}},
journal		= {Nature Communications},
year		= {2021},
volume		= {12},
number		= {1},
pages		= {1--14},
doi		= {10.1038/s41467-021-23342-2}
}

@Misc{0,
key		= {Teensy 4.0},
title		= {Teensy\textsuperscript{\textregistered} 4.0 Development
		  Board},
howpublished	= {PJRC website},
url		= {https://www.pjrc.com/store/teensy40.html}
}

@article{Rajendra_etal2006,
  title={Heart rate variability: a review},
  author={Rajendra Acharya, U and Paul Joseph, K and Kannathal, Natarajan and Lim, Choo Min and Suri, Jasjit S},
  journal={Medical and biological engineering and computing},
  volume={44},
  pages={1031--1051},
  year={2006},
  publisher={Springer}
}

@inproceedings{Narayanan_etal23,
  title={SPAIC: A sub-$\mu$W/Channel, 16-Channel General-Purpose Event-Based Analog Front-End with Dual-Mode Encoders},
  author={Narayanan, Shyam and Cartiglia, Matteo and Rubino, Arianna and Lego, Charles and Frenkel, Charlotte and Indiveri, Giacomo},
  booktitle={2023 IEEE Biomedical Circuits and Systems Conference (BioCAS)},
  pages={1--5},
  year={2023},
  organization={IEEE}
}

@article{Davidoff_etal23,
  title={Unraveling the relationship between heart rate and agitation in people with dementia},
  author={Davidoff, Hannah and Van Helleputte, Nick and Vandenbulcke, Mathieu and De Vos, Maarten and Hoof, Chris Van and Bossche, Maarten Van Den},
  journal={Alzheimer's \& Dementia},
  volume={19},
  pages={e078916},
  year={2023},
  publisher={Wiley Online Library}
}

@article{Vitale_etal22,
  title={Neuromorphic edge computing for biomedical applications: Gesture classification using emg signals},
  author={Vitale, Antonio and Donati, Elisa and Germann, Roger and Magno, Michele},
  journal={IEEE Sensors Journal},
  volume={22},
  number={20},
  pages={19490--19499},
  year={2022},
  publisher={IEEE}
}

@article{Donati_Indiveri23,
  title={Neuromorphic bioelectronic medicine for nervous system interfaces: from neural computational primitives to medical applications},
  author={Donati, Elisa and Indiveri, Giacomo},
  journal={Progress in Biomedical Engineering},
  volume={5},
  number={1},
  pages={013002},
  year={2023},
  publisher={IOP Publishing}
}

@article{Jarchi_etal16,
  title={Description of a database containing wrist PPG signals recorded during physical exercise with both accelerometer and gyroscope measures of motion},
  author={Jarchi, Delaram and Casson, Alexander J},
  journal={Data},
  volume={2},
  number={1},
  pages={1},
  year={2016},
  publisher={MDPI}
}

@article{Zanghieri_etal24,
  title={Event-based Estimation of Hand Forces from High-Density Surface EMG on a Parallel Ultra-Low-Power Microcontroller},
  author={Zanghieri, Marcello and Rapa, Pierangelo Maria and Orlandi, Mattia and Donati, Elisa and Benini, Luca and Benatti, Simone},
  journal={IEEE Sensors Journal},
  year={2024},
  publisher={IEEE}
}

@inproceedings{Sava_etal23,
  title={Feed-forward and recurrent inhibition for compressing and classifying high dynamic range biosignals in spiking neural network architectures},
  author={Sava, Rachel and Donati, Elisa and Indiveri, Giacomo},
  booktitle={2023 IEEE Biomedical Circuits and Systems Conference (BioCAS)},
  pages={1--5},
  year={2023},
  organization={IEEE}
}

@inproceedings{Zanghieri_etal23,
  title={Event-based Low-Power and Low-Latency Regression Method for Hand Kinematics from Surface EMG},
  author={Zanghieri, Marcello and Benatti, Simone and Benini, Luca and Donati, Elisa},
  booktitle={2023 9th International Workshop on Advances in Sensors and Interfaces (IWASI)},
  pages={293--298},
  year={2023},
  organization={IEEE}
}

@article{Zhu_etal22,
  title={A fitness training optimization system based on heart rate prediction under different activities},
  author={Zhu, Zetao and Li, Huining and Xiao, Jian and Xu, Wenyao and Huang, Ming-Chun},
  journal={Methods},
  volume={205},
  pages={89--96},
  year={2022},
  publisher={Elsevier}
}

@article{Heidenreich_etal11,
  title={Forecasting the future of cardiovascular disease in the United States: a policy statement from the American Heart Association},
  author={Heidenreich, Paul A and Trogdon, Justin G and Khavjou, Olga A and Butler, Javed and Dracup, Kathleen and Ezekowitz, Michael D and Finkelstein, Eric Andrew and Hong, Yuling and Johnston, S Claiborne and Khera, Amit and others},
  journal={Circulation},
  volume={123},
  number={8},
  pages={933--944},
  year={2011},
  publisher={Am Heart Assoc}
}

@article{Liu_etal23,
  title={Heart rate variability and risk of agitation in Alzheimer’s disease: the Atherosclerosis Risk in Communities Study},
  author={Liu, Kathy Y and Whitsel, Eric A and Heiss, Gerardo and Palta, Priya and Reeves, Suzanne and Lin, Feng V and Mather, Mara and Roiser, Jonathan P and Howard, Robert},
  journal={Brain Communications},
  volume={5},
  number={6},
  pages={fcad269},
  year={2023},
  publisher={Oxford University Press US}
}

@article{Caspar_etal17,
  title={Nonpharmacological management of behavioral and psychological symptoms of dementia: what works, in what circumstances, and why?},
  author={Caspar, Sienna and Davis, Erin D and Douziech, Aimee and Scott, David R},
  journal={Innovation in Aging},
  volume={1},
  number={3},
  pages={igy001},
  year={2017},
  publisher={Oxford University Press US}
}

@article{Zhao_etal23,
	title = {Learning inverse kinematics using neural computational primitives on neuromorphic hardware},
	volume = {1},
	issn = {2731-4278},
	url = {https://doi.org/10.1038/s44182-023-00001-w},
	doi = {10.1038/s44182-023-00001-w},
	number = {1},
	journal = {npj Robotics},
	author = {Zhao, Jingyue and Monforte, Marco and Indiveri, Giacomo and Bartolozzi, Chiara and Donati, Elisa},
	month = oct,
	year = {2023},
	pages = {1},
}

\end{document}